\documentclass[aps,amsmath,notitlepage,twocolumn,amssymb,prb,longbibliography]{revtex4-1}

\usepackage{graphicx}
\usepackage{dcolumn}
\usepackage{bm}
\usepackage[colorlinks=true,linkcolor = blue,citecolor=blue,urlcolor=blue,anchorcolor = blue]{hyperref}
\usepackage{wasysym}
\usepackage{stmaryrd}
\usepackage{verbatim}
\usepackage{subfigure}
\usepackage{amsmath}
\usepackage{times}
\usepackage[version=4]{mhchem}
\usepackage{braket}
\usepackage{siunitx}
\usepackage{siunitx}
\usepackage{amssymb}
\usepackage{mathrsfs}
\newcommand{\RNum}[1]{\uppercase\expandafter{\romannumeral #1\relax}}

\begin{document}

\title{\large \bf Effective Model for Rare-earth Kitaev Materials and its Classical Monte Carlo Simulation} 

\author{Mengjie Sun$^{1,2}$}
\thanks{These authors contributed to the work equally.}
\author{Huihang Lin$^{1}$}
\thanks{These authors contributed to the work equally.}
\author{Zheng Zhang$^{1,2}$}
\author{Yanzhen Cai$^{3}$}
\author{Wei Ren$^{3}$}
\author{Jing Kang$^{3}$}
\author{Jianting Ji$^{2}$}
\author{Feng Jin$^{2}$}
\author{Xiaoqun Wang$^{4}$}
\author{Rong Yu$^{1}$}
\author{Qingming Zhang$^{3,2}$}
\email[e-mail:]{qmzhang@ruc.edu.cn}
\author{Zhengxin Liu$^{1}$}
\email[e-mail:]{liuzxphys@ruc.edu.cn}

\affiliation{$^{1}$Department of Physics, Renmin University of China, Beijing 100872, China}
\affiliation{$^{2}$Beijing National Laboratory for Condensed Matter Physics, Institute of Physics, Chinese Academy of Sciences, Beijing 100190, China}
\affiliation{$^{3}$School of Physical Science and Technology, Lanzhou University, Lanzhou 730000, China}
\affiliation{$^{4}$Key Laboratory of Artificial Structures and Quantum Control of MOE, Shenyang National Laboratory for Materials Science, Shenyang 110016 and School of Physics and Astronomy, Tsung-Dao Lee Institute, Shanghai Jiao Tong University, Shanghai 200240, China}

\date{\today}

\begin{abstract}
Recently, the family of rare-earth chalcohalides were proposed as candidate compounds to realize the Kitaev spin liquid (KSL)\cite{JiantingJi:47502}. In the present work, we firstly propose an effective spin Hamiltonian consistents with the symmetry group of the crystal structure. Then we apply classical Monte Carlo simulations to preliminarily study the model and establish a phase diagram. When approaching to the low temperature limit, several magnetic long range orders are observed, including the stripe, the zigzag, the antiferromagnetic (AFM), the ferromagnetic (FM), the incommensurate spiral (IS), the Multi-$\pmb {Q}$ and the 120°. We further calculate the thermodynamic properties of the system,  such as the temperature dependence of the magnetic susceptibility and the heat capacity. The ordering transition temperatures reflected in the two quantities agree with each other. For most interaction regions, the system is magnetically more susceptible in the $ab$-plane than in the $c$-direction. The stripe phase is special, where the susceptibility is fairly isotropic in the whole temperature region. These features provide useful information to understand the magnetic properties of related materials. 

\end{abstract}

\maketitle

\section{INTRODUCTION}

Quantum spin liquid (QSL) are phases of matter beyond the Landau paradigm, which exhibits a disordered state even at absolute zero temperature due to strong quantum fluctuations. Long-range quantum entanglement instead of long-range correlations (i.e. long-range magnetic orders) are established in the QSL ground states. The elementary excitations in a gapped QSL obey fractional Abelian statistics, or even non-Abelian statistics\cite{Stern2010}. Candidate QSL materials with weak spin-orbit couplings have been found in triangular lattice, Kagome lattice and three-dimensional hyper-Kagome lattice \cite{PhysRevLett.91.107001, Itou2010, Han2012, PhysRevLett.109.067201, PhysRevB.78.094403, PhysRevLett.100.227201, PhysRevB.81.174417}. 

The Kitaev Spin Liquids (KSL), either gapless or gapped, are a special type of QSLs which are exact ground states of a simple honeycomb lattice spin model\cite{Kitaev2006}. A magnetic field can drive the gapless KSL into a non-Abelian chiral QSL which host non-Ablian anyons and have potential applications in quantum computations\cite{PhysRevLett.105.027204,Nasu2016,Barkeshli722}. Several materials, including the well studied $\alpha$-RuCl$_3$, were proposed to realize the Kitaev interactions\cite{PhysRevLett.102.256403, PhysRevLett.108.127204, PhysRevB.90.041112, PhysRevB.83.220403, PhysRevLett.119.037201, PhysRevLett.114.077202, PhysRevB.93.195158, PhysRevLett.122.047202, PhysRevB.92.235119,     PhysRevLett.120.187201, PhysRevLett.119.037201, PhysRevLett.119.227208, PhysRevB.96.041405, Yadav2016, Banerjee2018,PhysRevB.101.085120,PhysRevLett.118.107203,PhysRevB.101.045419}.  
The Ru$^{3+}$ ions behave like spin-1/2 spins whose exchange interactions contain the Kitaev terms\cite{PhysRevLett.112.077204, PhysRevLett.102.017205, PhysRevLett.114.096403, Banerjee2016,PhysRevLett.118.107203,Wang2017}. However, the ground state falls outside the KSL phase since the material exhibits zigzag-type long-range magnetic order at low temperatures\cite{PhysRevB.92.235119, PhysRevB.83.220403, PhysRevLett.102.256403, PhysRevLett.108.127204, PhysRevB.90.041112, PhysRevLett.119.037201}. This indicates that additional non-Kitaev ($K$) interactions exist, such as Heisenberg ($J$) terms and oﬀ-diagonal symmetric interactions (such as $\Gamma, \Gamma'$ terms) \cite{PhysRevLett.112.077204, PhysRevLett.118.107203, Laurell2020, PhysRevLett.120.187201, PhysRevLett.123.197201}. 

Besides the transition metal compounds, the rare-earth chalcogenide families are also QSL candidates. The rare-earth chalcogenide materials ARECh$_2$ (where A=alkali or monovalent$ $ ions, RE=rare-earth,Ch=O,S,Se) on triangular lattice, such as NaYbO$_2$, have attracted much attentions\cite{Liu_2018, zhang2020effective, PhysRevB.103.035144}. On the other hand, the family of honeycomb lattice rare-earth chalcohalides REChX (RE=rare earth, Ch=O,S,Se,Te, X=F,Cl,Br,I), such as  YbOCl and SmSI, provides alternative possibilities to realize Kitaev QSLs\cite{JiantingJi:47502}.The REChX family is a good quasi-two-dimensional material due to the lager distance between layers. Furthermore, the strong crystalline electronic field (CEF) owing to chalcogen ion ensures that the effective spin of the rare earth ion is spin-1/2. The strong spin-orbital coupling results in anisotropic exchange interactions. The physical properties of YbOCl in the temperature range of $1.8 K\sim 300 K$ have been reported in Ref.\onlinecite{JiantingJi:47502}, but the microscopic effective model need to be investigated.

In this work, we first discuss possible interactions in YbOCl based on symmetry analysis. The potential Hamiltonian includes the Heisenberg terms($J$), the Kitaev terms($K$), the off diagonal terms ($\Gamma$,$\Gamma^{'}$) and the DM interactions. We then study the resultant model Hamiltonian using classical Monte Carlo simulation, and obtain the preliminary phase diagram,  which includes the stripe phase, the zigzag phase, the antiferromagnetic (AFM) phase, the ferromagnetic (FM) phase,  the 120° phase,  the incommensurate spiral (IS) phase and the Multi-$\pmb {Q}$ phase. We further calculate thermodynamic properties  such as the temperature dependence of the magnetic susceptibility and the heat capacity, which may help to determine the interaction parameters and to understand the magnetic and thermal properties at low temperatures in real materials.

\begin{figure*}[htb]
\centering
\includegraphics[height=4.cm,width=16cm]{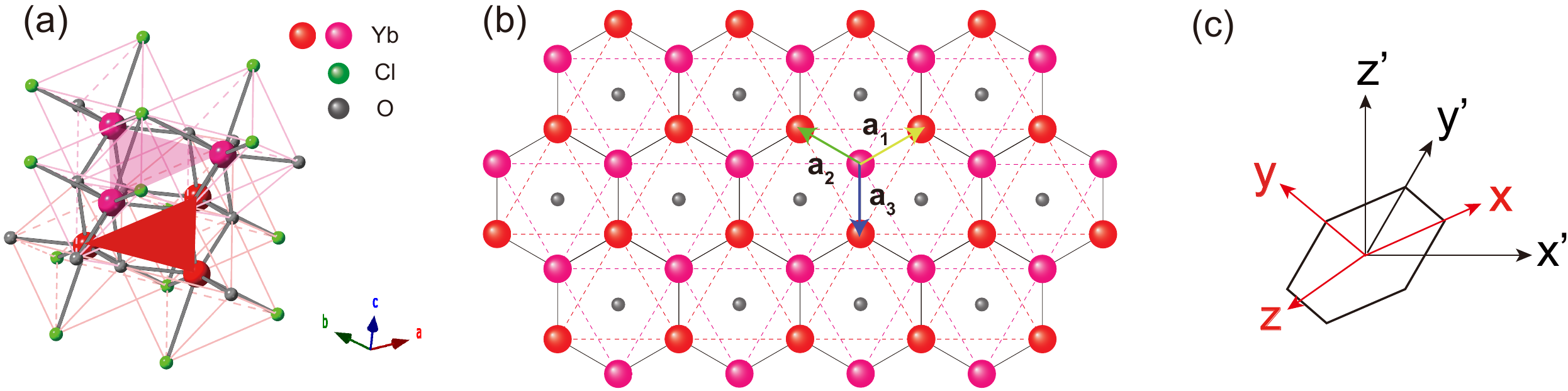}
\caption{ (a) The crystal structure of YbOCl. (b) The quasi honeycomb lattice `plane' (called the $ab$-plane with the normal direction $c$) formed by the Yb$^{3+}$. (c) The conventional frame ($\hat x', \hat y', \hat z'$) VS. the new frame ($\hat x, \hat y, \hat z$), where $\hat x'={1\over\sqrt2}(\hat x-\hat y), \hat y'={1\over\sqrt6}(\hat x+\hat y-2\hat z), \hat z'={1\over\sqrt3}(\hat x+\hat y+\hat z)$.}
\label{Fig.1}
\end{figure*}

\section{Symmetry allowed  spin-spin Interactions}

\subsection{Structure and Symmetry}

The layered rare-earth material YbOCl belongs to the SmSI family with ABCABC stacking structure, and its space group symmetry is 166 (R$\overline{3}m$) \cite{Song1994,JiantingJi:47502}, as illustrated in Fig.\ref{Fig.1}(a). The Yb$^{3+}$ ion form two adjacent layers of triangular lattice, whose lattice constant in each layer reads $d_2$=3.72 Å. The two layers nest with each other to form a rough `honeycomb lattice', where the distance between nearest neighbors (the nearest bond link the two triangular layers) is $d_1$ = 3.545 Å, which is smaller than $d_2$.  Although the two layers of triangular lattice do not have $C_6$ rotation or $\sigma_h$ mirror reflection symmetry, the combined operation $S_6=C_6\sigma_h$ is indeed a symmetry element (here the mirror plane $\sigma_h$ locates at the middle of the two triangular layers). Moreover, the distance between two adjacent `honeycomb layers' is $d_3$ = 6.443 Å, which is much larger than $d_2$. Therefore, we can treat YbOCl as a quasi-two-dimensional material with honeycomb lattice structure. 

Each Yb$^{3+}$ ion is surrounded by three Cl$^-$ (the angle formed by the bonds Cl-Yb-Cl is 85.505°) and four O$^{2-}$ (the angle formed by the bonds O-Yb-O is 76.229° or 114.52°). These anions form the complex polyhedral CEF environment for the magnetic ion Yb$^{3+}$ with site group $\mathscr C_{3v}$. The electronegativity of coordination anion Cl$^-$ and O$^{2-}$ are very strong, and the energy splitting caused by the CEF is of order of $65$mev, which is much larger than the band width ($\sim$10mev) of the low-lying energy level. Therefore, the system maintains an effective spin $S = 1/2$ local moment at a large temperature range. The nearest neighbor sites perform super-exchange interaction through O$^{2-}$, while the next nearest-neighbor sites perform it through both Cl$^{-}$ and O$^{2-}$. Since the orbits of the Yb$^{3+}$ ions are fairly local, the intensity of the super-exchange interaction is relatively weak which is of order of 1 K\cite{zhang2020effective,PhysRevLett.115.167203,Li2015,PhysRevLett.117.097201,Shen2016,PhysRevLett.117.267202,PhysRevLett.118.107202,PhysRevB.95.165110}.


Since the R$\bar{3}$m group is symmorphic, its point group $D_{3d}$ precisely describes the symmetry of the center of the unit cell. The $D_{3d}$ point group is generated by $S_6$ and $C_2$, where the $S_6$ symmetry operation was illustrated previously. The two-fold rotation axis $C_2$ lies in the mirror plane of $S_6$ and points from the center of the unit cell to the bond center of two adjacent Yb$^{3+}$ ions (this $C_2$ axis is perpendicular to the bond direction, in contrast to the standard orientation of the Kitaev model where the $C_2$ axis is along the bond direction \cite{PhysRevB.84.024420, PhysRevB.86.085145}). The $D_{3d}$ point group determines the possible effective interactions between the Yb$^{3+}$ spins. 

The $4f$ electrons in Yb$^{3+}$ ions have a strong spin-orbit coupling ( of order of 0.36eV). This indicates that the effective interactions have a strong anisotropy. Actually, the strong electronegativity of O$^{2-}$ makes the bonds strongly ionic, which causes a higher tendency of performing super-exchange interactions between cations and coordination anions. Meanwhile, adjacent polyhedrons share the same side, thus the Heisenberg interactions are suppressed and the rest interactions are strongly anisotropic\cite{PhysRevLett.102.017205}.

Recall that the Kitaev honeycomb lattice model also has a $D_{3d}$ point group symmetry when considering spin-orbit coupling, which is the same as the point group of YbOCl. Therefore it is natural to infer that the low-energy effective model of YbOCl contains the Kitaev type interactions (the $K$ terms). This makes YbOCl a possible Kitaev material. Like most Kitaev materials, the Kitaev interaction has negative sign($K<0$).  Other interactions are also allowed as long as they are consistent with the $D_{3d}$ symmetry. We will discuss the effective spin-spin interactions in subsection \ref{sec:H}. 

\begin{figure*}[t]
\centering
\includegraphics[height=6.5cm,width=18cm]{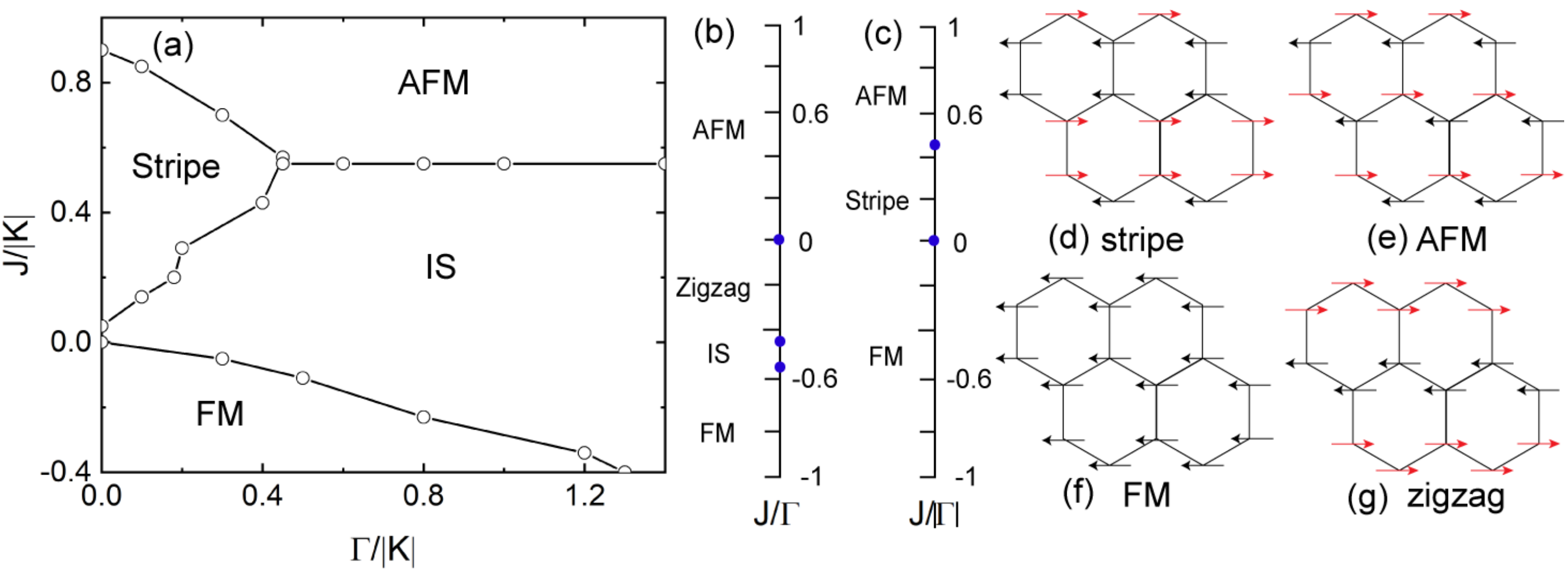}
\caption{(a) Phase diagram of the classical $J$-$K$-$\Gamma$($K$ $<$ 0) model, which contains the stripe, the antiferromagnetic (AFM), the ferromagnetic (FM) and the incommensurate spiral (IS) phases. (b) Phase diagram of the classical $J$-$\Gamma$ model for $\Gamma$ $>$ 0; the transitions occur at $J$/$\Gamma$ = 0,-0.45 and -0.55. (c) Phase diagram of the classical $J$-$\Gamma$ model for $\Gamma$ $<$ 0; the transitions occur at $J$/$|\Gamma|$ = 0 and 0.45. (d)-(g) show the spin configurations in different magnetic orders. Except for the IS phase, all the other magnetic orders are collinear. The orientation of the magnetic momentums in each phase is the following: the stripe $\pmb M_i\parallel [1,0,0]$, the AFM $\pmb M_i\parallel[1,1,1]$, the FM $\pmb M_i\parallel[5,3,-8]$, the zigzag $\pmb M_i\parallel[1,-1,1]$. }
\label{Fig.2}
\end{figure*}

\subsection{Low-energy effective Hamiltonian Model}\label{sec:H}


As mentioned above, the symmetry group $D_{3d}$ of YbOCl is consistent with that of the Kitaev model. In the Kitaev model, owing to spin-orbit coupling, the $c$-axis is parallel to the ${1\over\sqrt3}(\hat x+\hat y+\hat z)$ direction in the spin frame such that a $C_3$ rotation permutes $\hat x, \hat y, \hat z$ cyclically. Here we adopt the same convention. Furthermore, the horizontal direction (i.e. the $C_2$ axis) is identified as ${1\over\sqrt2}(\hat x-\hat y)$, and the vertical direction is indentified as ${1\over\sqrt6}(\hat x+\hat y-2\hat z)$. Hence we obtain the most general Hamiltonian of the family of rare-earth chalcohalides \cite{10.21468/SciPostPhysCore.3.1.004,PhysRevX.1.021002,PhysRevLett.115.167203,PhysRevB.98.054408,PhysRevB.94.035107,zhang2020effective, PhysRevLett.115.167203,Li2015,PhysRevLett.117.097201,Shen2016,PhysRevLett.117.267202,PhysRevLett.118.107202,luo2020gapless,PhysRevB.97.214433}:

\begin{eqnarray}\label{Habc}
H=\sum_{\langle i,j\rangle \in \alpha \beta (\gamma)} &&KS^\gamma_iS^\gamma_j +J\pmb{S}_i\cdot\pmb{S}_j +\Gamma (S_i^\alpha S_j^\beta + S_i^\beta S_j^\alpha) \notag\\
&&+\Gamma^{'}(S_i^\alpha S_j^\gamma+S_i^\gamma S_j^\alpha+S_i^\beta S_j^\gamma+S_i^\gamma S_j^\beta) \notag\\
&&+\sum_{\langle\langle ij \rangle\rangle } \pmb D_{ij}\cdot(\pmb {S}_i\times\pmb{S}_j), 
\end{eqnarray}
where $\alpha$, $\beta$, $\gamma$ label the type of the nearest neighbor bonds and the spin indices, and $\langle i,j\rangle$,$\langle\langle i,j\rangle\rangle$ denote nearest-neighbor and next nearest-neighbor sites, respectively. 

The last term in (\ref{Habc}) is the DM interaction on next nearest-neighbor bonds, which is assumed to be stronger than other interactions on the same bonds. Structurally, the $\pmb D_{ij}$ vector takes the following form
$\pmb{D}_{ij} = D(\pmb{r}_i \times \pmb{r}_j$) with $\pmb{r}_i$ the position vector  pointing from the $i$th magnetic ion to the coordination anion which mediates the super-exchange. For YbOCl, the next nearest-neighbor rare-earth ions main exchange throng the O$^{2-}$ ions (the other type anions Cl$^{-}$ are far away from the next-nearest bonds and have weaker effects). Since the O$^{2-}$ anions are almost locating in the cation layer, it is expected that the DM vector $\pmb D_{ij}$ is almost pointing along the $c$-direction (namely the [1,1,1] direction in the spin frame).



Here we clarify the difference between the conventional spin frame (we label the axes as $\hat x', \hat y', \hat z'$) and the one introduced above. Since the $D_{3d}$ point group has only one high-symmetry axis $\hat c$, conventionally the $\hat c$ axis is chosen to be the $\hat z'$-axis, the horizontal line is identified with the $\hat x'$-axis and the vertical line is identified with the $\hat y'$ axis (see Fig.\ref{Fig.1} for illustration). In other words, in the conventional spin frame, the axes are parallel to the corresponding ones of the lattice frame. The spin operators in the two frames are related in the following way,
\begin{eqnarray*}
&&S^{'x} = {1\over\sqrt 2}(S^x-S^y),\\ 
&&S^{'y} = {1\over\sqrt 6}( S^x+S^y-2S^z),\\
&& S^{'z} = {1\over\sqrt 3}(S^x+S^y+S^z).\ 
\end{eqnarray*}

The symmetry operations act differently in the two frames. Owing to spin-orbital coupling, the point group symmetry elements not only transform the sites but also transform the spins in the same way.  For instance, if $i$ and $j$ label two sites on a nearest neighbor $z$-bond, then in the symmetric spin frame, $S_6$ transforms $S_i^{x,y,z}$ to $S_j^{z,x,y}$, and $C_2$ operation transforms $S_i^{x,y,z}$ to $-S_j^{y},-S_j^{x},-S_j^{z}$, respectively.  
However, in the conventional frame, $S_6$ transforms $S_i^{' x},S_i^{'y},S_i^{'z}$ to 
\begin{eqnarray*}
&&S_6 S_i^{' x} S_6^{-1} = -{1\over2}S_j^{'x}-{\sqrt3\over2}S_j^{'y},\\ 
&&S_6 S_i^{' y} S_6^{-1} = {\sqrt3\over2}S_j^{'x}-{1\over2}S_j^{'y}, \\
&&S_6 S_i^{' z} S_6^{-1} =  S_j^{'z}
\end{eqnarray*}
respectively, and $C_2$ transforms $S_i^{' x},S_i^{'y},S_i^{'z}$ to 
$$C_2 S_i^{' x} C_2^{-1} = S_j^{'x},\ C_2 S_i^{' y} C_2^{-1} =  -S_j^{'y},\ C_2 S_i^{' z} C_2^{-1} = -S_j^{'z}$$ respectively.

In the conventional spin frame, the model (\ref{Habc}) takes a different form [see appendix \ref{app:A}]. Under the above rule of symmetry transformations, the transformed model still preserves the $D_{3d}$ symmetry group.



\begin{figure*}[htb]
\centering
\includegraphics[height=7.8cm,width=14.5cm]{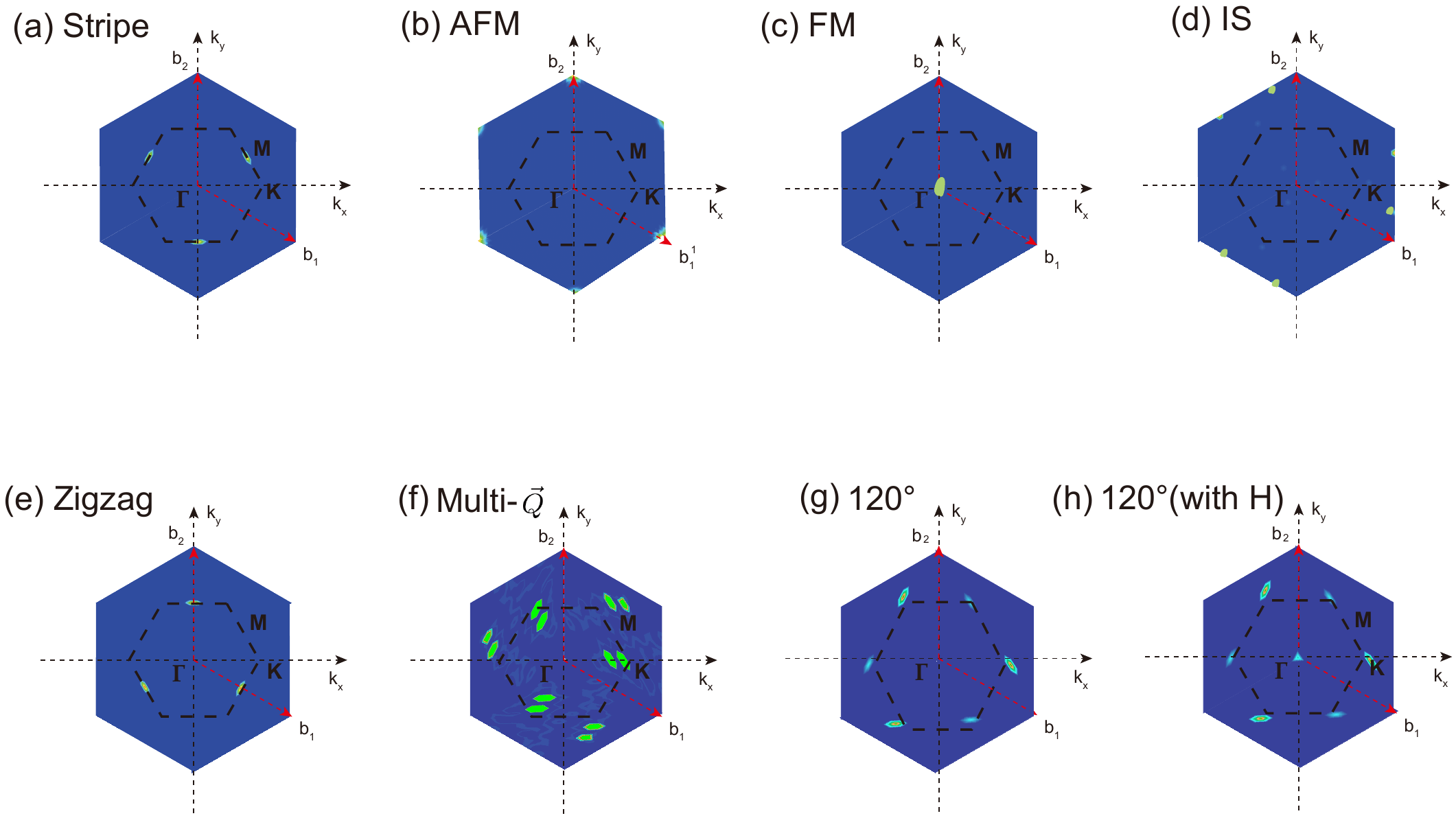}
\caption{Low-$T$ MC static structure factor for $J$-$K$-$\Gamma$-$\pmb D$ ($K$ $<$ 0) model. Bright spots indicate Bragg peaks. The dotted inner hexagon denotes the ﬁrst Brillouin zone. (a)$\sim$(h) respectively illustrate the static structure factor of the stripe, the AFM, the FM, the IS, the zigzag, the Multi-$\pmb {Q}$, the 120° order and the 120° order in magnetic field.}
\label{Fig.3}
\end{figure*}

\section{Preliminary Results of the Model}

\subsection{Classical Monte Carlo Simulation}
We use the classical Monte Carlo (MC) method to numerically simulate the model (\ref{Habc}), where the spins are treated as classical three-component unit vectors. Since $\Gamma'$ is generally small in Kitaev materials\cite{PhysRevB.93.214431,PhysRevB.97.134424}, we set $\Gamma'=0$ in our calculations. We firstly investigate the case with $\pmb D_{ij}=0$, and then study the effect of the DM interactions. We aim to determine the phase diagram of the classical ground states, and provide the critical temperatures of each ordered phase.

In our classical MC simulation, the spin configurations are distributed with a probability proportional to $\exp\{-{E\over kT}\}$ (where $E$ is the total energy of the system) according to Boltzmann statistics.  Metropolis algorithm is applied to update the spin configurations in the MC sampling.  

However, for the honeycomb lattice model the slowing down of configuration updating is very serious when approaching the critical temperatures. To solve this problem, we use the microcanonical over-relax algorithm to reduce the autocorrelation time and replace the usual MC steps by the `mixed' MC steps. Generally, the over-relax algorithm makes the spin rotate at any angle around the local equivalent field\cite{Kanki2005}, so that the total energy will not be changed. The simplest and most effective method is to make the spin rotate $\pi$ angle around the local equivalent field, which is actually adopt in our MC simulations. There is no need to generate random numbers to achieve such spin flip, thus saving the simulation time. Practically we extend each usual MC step (a usual MC step consists of $N$ times of trial spin flip) by a mixed MC step, which consists of a usual MC step and some over-relaxation updates. The deterministic change of spin is very important to reduce the autocorrelation time at low temperatures. According to the lattice size, we increase the number of over-relaxation updates per MC step. 

Finally, to avoiding being trapped into local minimum, we apply the annealing algorithm in the low temperature region.

In our simulation, the system size is $N=2\times L^2$, with $L=24$. Periodic boundary condition is adopt.  For each parameter point, the system is preheated by 5$\times 10^5$ times to reach the thermal equilibrium, then measurements are made every three MC steps. The total number of the times of measurements is 5$\times 10^5$.

\begin{figure*}[t]
\centering
\includegraphics[height=7cm,width=18cm]{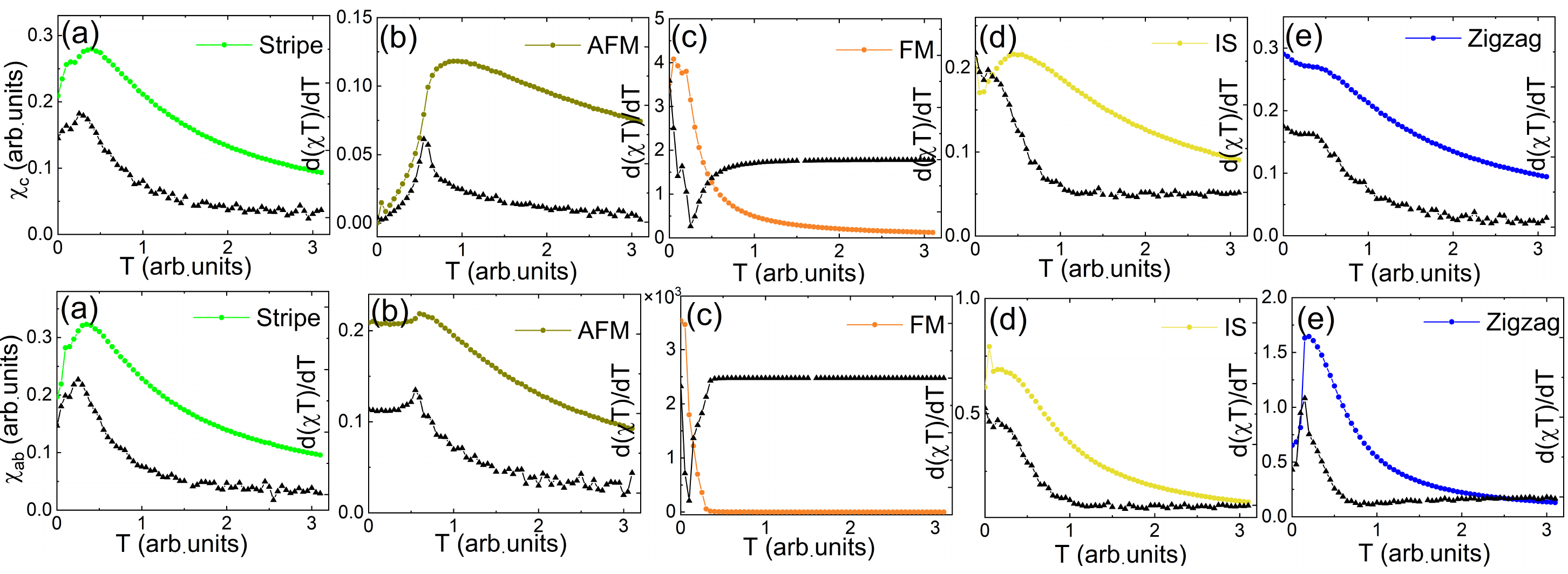}
\caption{The susceptibility $\chi_c$ and $\chi_{ab}$ for different magnetic ordered phases: (a) the Stripe with $\Gamma/|K|=0.1,J/|K|=0.5$, (b) the AFM with $\Gamma/|K|=0.8, J/|K|=1$, (c) the FM with $\Gamma/|K|=0.4, J/|K|=-0.35$, (d) the IS with $\Gamma/|K|=0.8, J/|K|=0.3$, (e) the Zigzag with $J/\Gamma=-0.2(K=0)$. The colored curves show the susceptibility $\chi$, the dark ones illustrate $d(\chi T)/dT$ (the data have been multiplied by some constant to guide the eyes) from which the transition points can be easily read out. }

\label{Fig.4}
\end{figure*}

\begin{figure*}[t]
\centering
\includegraphics[height=3.5cm,width=18cm]{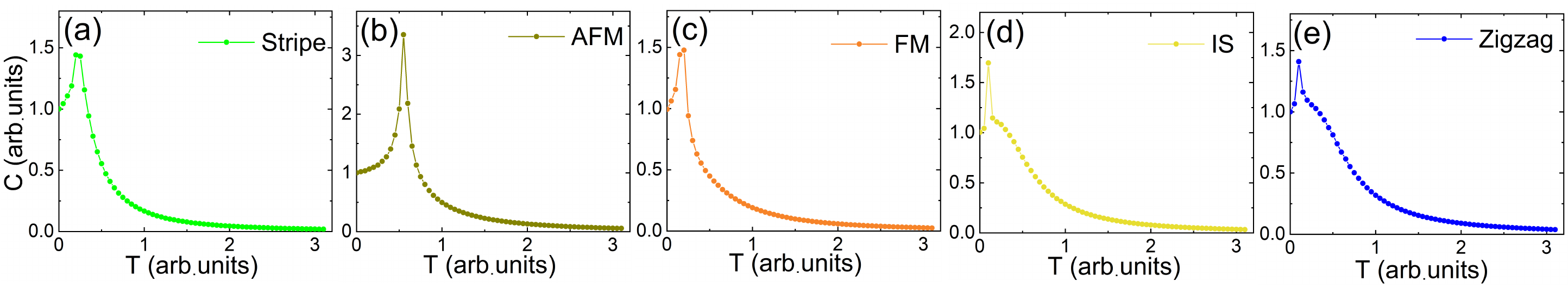}
\caption{The heat capacity of different magnetically ordered phases and their corresponding high temperature paramagnetic phases. The interaction parameters of (a)$\sim$(e) are the same as the corresponding ones provided in the caption of Fig.\ref{Fig.4}.}
\label{Fig.5}
\end{figure*}

\subsection{The Phase Diagram}

Fig.\ref{Fig.2} shows the magnetic phase diagram of the classical ground states with $\Gamma'=0, \pmb D_{ij}=0$ (the model with the same parameters was studied using different methods\cite{PhysRevLett.112.077204,PhysRevLett.123.197201,PhysRevB.97.075126,PhysRevB.96.064430,PhysRevB.98.060405,PhysRevB.95.024426,liu2020revealing,rao2021machinelearned,PhysRevB.103.054410}). Five phases are obtained, namely, the stripe, the antiferromagnetic (AFM), the ferromagnetic (FM), the incommensurate spiral (IS) and the zigzag.
Except for the IS phase, all the other magnetic orders are collinear. To identify the magnetic phases, we plot the spin configurations in real space, as shown in Fig.\ref{Fig.2} (d)$\sim$(g). The orientation of the magnetic momentums in each phase is the following: the stripe phase $\pmb M_i\parallel [1,0,0]$ (equivalent orientations related by the symmetry group also include $[0,1,0]$ and $[0,0,1]$, here and later we only list one of them), the AFM phase $\pmb M_i\parallel[1,1,1]$, the FM phase $\pmb M_i$ approximately parallel to $[5,3,-8]$, the zigzag phase $\pmb M_i\parallel[1,-1,1]$.

We also calculate the static spin structure factor. Each phase is characterized by the Bragg peaks in the reciprocal lattice, as shown in Fig.\ref{Fig.3}. The stripe phase and zigzag phase are both peaked at three $C_3$ symmetry-related $M$ points, namely the midpoint of the hexagonal edge of Brillouin zone. Since the two types of orders have the same inter unit-cell patterns but different intra unit-cell patterns, the positions of the Bragg peaks in the two phases differ by a reciprocal lattice vector. Similarly, the FM phase has a Bragg peak at the center of the BZ, while the AFM phase is peaked at the boundary of the BZ. The static structure factor of the IS phase is peaked at several equivalent non-high-symmetry points which are related by symmetry.

The magnetic ground state of the system is the stripe phase when 0 $< J/|K| <$ 0.9 and is the AFM phase when $J/|K| >0.9$, where $J$ is antiferromagnetic interaction and $\Gamma$=0. When the size of $J$ and $K$ are basically the same, the Heisenberg ($J$) interaction dominates and the system enters the AFM phase. When both $J$ and $K$ are ferromagnetic, the ground state of the system naturally enters the FM phase. Next, we focus on the regulation of  $\Gamma$ in the pure $J$-$K$ model. Antiferromagnetic $\Gamma$ will cause transition from the stripe phase to the AFM phase or the IS phase.  And $\Gamma$ will also cause transition from the FM phase to the IS phase. Experimentally, most of the Kitaev materials fall in the zigzag phase or the IS phase, indicating the existence of $\Gamma$ and other interactions. 

\par

In order to analyze the regulation of $\Gamma$ on $J$ interaction, we simulate the ground state magnetic phase diagram of  $J/|\Gamma|$ when $K$ = 0, as shown in Fig.\ref{Fig.2}(b) and Fig.\ref{Fig.2}(c). $\Gamma$ in Fig.\ref{Fig.2}(b) is the antiferromagnetic exchange interaction. When $J/\Gamma>$ 0, the ground state of the system is the AFM phase. When $J$ = 0, the system enters the multi-$\pmb{Q}$ state.
When -0.45 $<$ $J/\Gamma$   $<$ 0, the system transits to the zigzag phase, namely, the system can also enter the zigzag phase when only $J$ and $\Gamma$ interaction exist. When -0.55 $<$ $J/\Gamma$  $<$ -0.45, the system enters the IS phase again. It can be seen that the zigzag phase and the IS phase are always related to each other, indicating that they may have similar energy. When  $J/\Gamma$   $<$ -0.55, the system enters the conventional FM phase.

Fig.\ref{Fig.2}(c) is the the phase diagram with $\Gamma<0, K=0$. When $J$ is antiferromagnetic interaction and $J/|\Gamma| >$ 0.45, the ground state of the system is the AFM phase, which is natural under large antiferromagnetic Heisenberg interaction. But when 0 $<J/ |\Gamma|<$ 0.45, the system enters the stripe phase.  When $J$ = 0, the system also enters the multi-$\pmb {Q}$ state.  And when $J/|\Gamma|$   $<$ 0, the system no longer enters the strange magnetic ordered phase, but directly enters the FM phase. 
 
 \begin{figure*}[t]
\centering
\includegraphics[height=5.8cm,width=18cm]{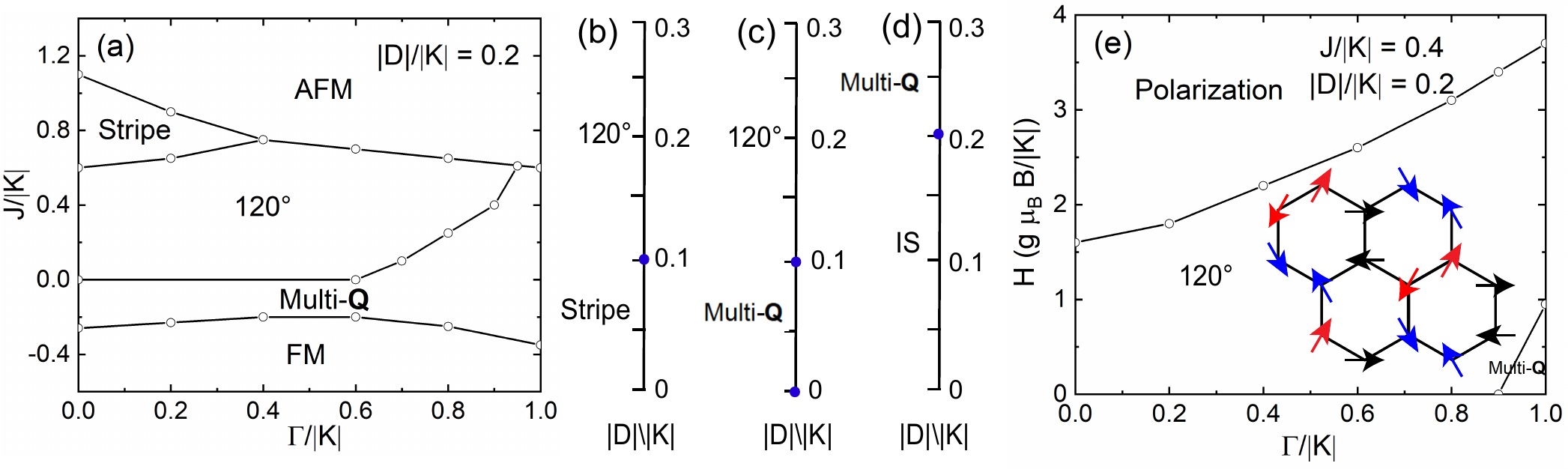}
\caption{(a) Phase diagram of the classical $J$-$K$-$\Gamma$($K$ $<$ 0, $|D|/|K|$=0.2) model, which contains the stripe, the antiferromagnetic (AFM), the ferromagnetic (FM), the 120° order and the Multi-$\pmb {Q}$ phases. (b) For $J/|K|=0.3$, $\Gamma/|K|=0.1$,  the transitions occur at $|D|/|K|$=0.1. (c) For $J/|K|=0.2$, $\Gamma/|K|=0.5$,  the transitions occur at $|D|/|K|$=0 and 0.1. (d) For $J/|K|=-0.1$, $\Gamma/|K|=0.8$, the transitions occur at $|D|/|K|$=0.2. (e) Phase diagram of the $\Gamma/|K|$ with $\pmb {H}\parallel [1,1,1]$ ($J/|K|$=0.4 and $|D|/|K|$=0.2); the inset illustrates the in-plane spin configurations in the 120° ordered phase. }
\label{Fig.6}
\end{figure*}

\subsection{The Thermodynamic Quantities}

In order to study the magnetic properties of each magnetic ordered phase, we simulate the thermodynamic quantities, including the susceptibility and the heat capacity, for each phase in the phase diagram. 

From the fluctuation-dissipation theorem, above thermal quantities can be evaluated in the MC simulations from the correlation functions of the corresponding physical quantities, namely,

\begin{eqnarray*}
&&C(T)=(\langle E^2\rangle-\langle E\rangle^2)/(NT^2),\\ 
&&\chi_\alpha(T)=(\langle M_\alpha^2\rangle -\langle M_\alpha\rangle ^2)/NT.
\end{eqnarray*}

The susceptibility, as a tensor, has the same point group symmetry as the crystal, which is $D_{3d}$. Since the material is uni-axial, so the magnetic susceptibility has two different eigenvalues $\chi_c$ and $\chi_{ab}$ (the symmetry guarantees that the magnetic susceptibility in the $ab$-plane is isotropic). In all of the magnetic phases, $\chi_{c}$ is smaller than $\chi_{ab}$, meaning that the spins are more susceptible along the $ab$-plane and the $c$-direction is a hard axes. At high temperature region, the susceptibility obeys the Curie law. In order to accurately identify the phase transition temperatures,  the quantity $d(\chi T)/dT$ is also plotted, whose singular points indicate the transitions.

For most of the ordered phases,  the susceptibility shows a $\lambda$-shape peak near the critical point in two directions. However, for the zigzag phase there is no obvious peak in $\chi_c$ near the phase transition point. Instead, a plateau appears near the phase transition point and then monotonously increases with decreasing temperature. $\chi_{ab}$ has obvious $\lambda$-shape peak,  and the phase transition occurs at the point where the susceptibility decreases most rapidly.  

The $\chi_{ab}$ of IS is also very unusual, which shows a strange double-peak structure, the one at the low-temperature side is sharp and the other is relatively boarder. The structure factor and susceptibility demonstrate that the sharp peak represents the phase transition. Therefore, the phase transition temperature of IS order is very low, indicating that the system remains fluctuating and forms the IS order only at very low temperatures.

The magnetic susceptibility in the stripe phase is almost isotropy. The AFM order is arranged in the [1,1,1] direction, so $\chi_c$ approaches zero but $\chi_{ab}$ remains finite at zero temperature.

The transition temperatures indicated from the heat capacity (see Fig.\ref{Fig.5}) are basically consistent with those obtained from the $d(\chi T)/dT$. The heat capacity shows a sharp peak in the transition point to the stripe phase, or the AFM phase, or the FM phase.The zigzag and IS phase are special, since there is a shoulder like structure in the heat capacity at the right hand side of the transition point.

These features of the magnetic susceptibility in Fig.\ref{Fig.4} and the specific heat in Fig.\ref{Fig.5} are helpful to identity the magnetic orders in real materials.

However, it should be cautioned that in the low temperature limit, quantum effects can not be neglected even in the ordered phases. In the pure classical model, since the energy is a continuous function of spin configurations, the specific heat has a finite residue value at zero temperature (it is of order 1 in all of the magnetic phases, see Figure \ref{Fig.5}). By considering the quantum corrections in the semi-classical linear spin wave theory, the residue specific heat should be zero. If the magnon excitations are gapless, then the specific heat should decay to zero in power law when $T$ approaches to zero, namely $C(T)\propto T^n$ (where $n$ is an integer) ; otherwise, if the magnons are gaped, then $C(T)$ decays exponentially with lowering temperature, namely $C(T)\propto e^{-\Delta/k_BT}$ with $\Delta$ is the magnon gap and $k_B$ is the Boltzmann constant. In both cases, the $C=0$ at $T=0$, which is consistent with the zeroth law of thermodynamics. Similarly, the low-temperature behavior of the magnetic susceptibility in Fig.\ref{Fig.4} should also be strongly affected by quantum fluctuations. 
 
\subsection{Effection Of The DM Interaction}

In this subsection we study the effect of the next nearest-neighbor DM interaction with $\pmb D\parallel  [1,1,1]$. A finite $|D|$  dramatically changes the phase diagram. For instance, at fixed $|D|/|K|$=0.2, the magnetic phase diagram of the classical ground states is shown in Fig.\ref{Fig.6}(a). 

Compared to Fig.\ref{Fig.2}(a), the main difference is that the IS phase is replaced by two new phases (a shortened $IS$ phase should remains if $|D|$ is smaller), the 120$^\circ$ phase and the multi-$\pmb Q$ phase. The static structure factor of the 120° phase is peaked at $K$ points while the static structure factor of the multi-$\pmb Q$ phase is peaked more than one non-equivalent momentum points, see Fig.\ref{Fig.3}(g) and (f), respectively.  

The 120$^\circ$ phase occurs at the region $0<J<0.6$ and $\Gamma/|K|<0.94$. The multi-$\pmb Q$ phase roughly locates at $-0.2<J<0$ as $\Gamma/|K|<0.6$. When $\Gamma/|K|>0.6$, the multi-$\pmb Q$ phase competes with the 120$^\circ$ phase and completely beats it as $\Gamma/|K|>0.94$. Furthermore, the sizes for all of the rest phases are shortened.  The lower boundary of the AFM phase slightly goes up, and the upper boundary of the FM phase moves down. The stripe phase as a whole slightly moves up, giving part of its place to the 120$^\circ$ phase.


From above analysis, we can see the strong preference of the 120$^\circ$ phase and the multi-$\pmb Q$ phase by the out-of-plane DM interactions. In order to explore the critical DM interaction, we simulate the magnetic phase transitions with increasing $|D|/|K|$ in different magnetic phase regions, as shown in Fig.\ref{Fig.6}(b)-(d). 

At $\Gamma/|K|=0.1, J/|K|=0.3$, the stripe phase turns to the 120° phase at $|D|/|K|=0.1$. 

At $\Gamma/|K|=0.5, J/|K|=0.2$, the IS ($J$ is antiferromagnetic interaction) phase is replaced by the Multi-$\pmb {Q}$ phase at an extremely small DM interaction, and then transits to the 120° phase at $|D|/|K|=0.1$. 

At $\Gamma/|K|=0.8, J/|K|=-0.1$, the IS phase survives for a finite region of $D$, until it replaced by the Multi-$\pmb {Q}$ phase as $|D|/|K|>0.2$.


Finally, we preliminarily study the effect of magnetic field in the presence of DM interactions. Fig.\ref{Fig.6}(e) shows the phase diagram of the $\Gamma/|K|$ with $\pmb {H}\parallel[1,1,1]$, $J/|K|$=0.4 and $|D|/|K|$=0.2. The 120° phase is robust against magnetic field with its static structure factor peaks at both the $K$ points and the center of the BZ [see Fig.\ref{Fig.3}(e) for illustration]. The inset illustrate the in-plane ordering pattern (the out-of-plane components which are partially polarized by the field are not shown). Firstly, when the field is weaker than $g\mu_B H/|K|$, the  Multi-$\pmb {Q}$ phase is suppressed and the 120° phase is enlarged. Secondly, when the filed is strong enough, the 120° order will be destroyed at a critical field strength. It turns out that the critical field strength increases with the increase of $\Gamma/|K|$. 

When quantum effects are taken into account, an out-of-plane DM interaction and a magnetic field can cause thermal Hall effect since the magnon excitations (in the ordered phase) or the spinon excitations (in the disordered phase) will feel nonzero Berry phase\cite{doi:10.1021/acs.nanolett.0c00363, Gao2020}. Furthermore, in-plane DM interactions and out-of-plane magnetic field may generate Skyrmion excitations at suitable temperature region\cite{PhysRevB.100.245106}. These possibilities make the rare-earth chalcohalide a platform to explore the interesting physics. 


\section{Conclusions and discussions}

In summary, we have studied the magnetism of rare-earth chalcohalides as candidate materials of Kitaev QSL. Based on the crystal structure and symmetry group, a low-energy effective model is proposed. We calculate the classical phase diagram of the magnetic ground state of the $J$-$K$-$\Gamma$-$\pmb D$ model using classical Monte Carlo method. We identify the classical magnetic orders in the ground states, including the stripe,  the zigzag, the antiferromagnetic (AFM), the ferromagnetic (FM), the incommensurate spiral (IS), the multi-$\pmb Q$ and the 120$^\circ$ order. The next nearest-neighbor DM interaction strongly prefers the 120$^\circ$ and the multi-$\pmb Q$ phase. The temperature dependence of the magnetic susceptibility and the heat capacity are provided, which may help to experimentally identify the magnetic orders and to understand their low-temperature behaviors. 



The family of rare-earth chalcohalides provides an ideal platform for the further study of Kitaev QSLs. The parameters of the effective model for the YbOCl and other materials in the family need to be determined from further experimental measurements(such as inelastic neutron scattering). On the other hand, the quantum phase diagram of the model with nonzero next-nearest neighbor interactions (DM and other possible interactions) needs to be figure out using quantum many-body computation methods. We leave these studies to future work.

\section*{ACKNOWLEDGEMENTS}
This work was supported by the National Key Research and Development Program of China (Grants No. 2017YFA0302904 and No. 2016YFA0300504), the NSF of China (Grants No. U1932215 and No. 11774419), and the Strategic Priority Research Program of the Chinese Academy of Sciences (Grant No. XDB33010100). Q.M.Z. acknowledges the support from Users with Excellence Program of Hefei Science Center and High Magnetic Field Facility, CAS. Z.X.L. is supported by the Ministry of Science and Technol- ogy of China (Grant No. 2016YFA0300504), the NSF of China (Grants No.11574392 and No. 11974421), and the Fundamental Research Funds for the Central Uni- versities and the Research Funds of Renmin University of China (Grant No. 19XNLG11).

\appendix
\section{Hamiltonian transformation in different coordinate systems}\label{app:A}

Notice that the $D_{3d}$ point group has only one high-symmetry axis. Usually this axis is chosen to be the $\hat{z}^{'}$-axis for both the lattice frame and the spin frame. The Hamiltonian will be complicated when the spin and lattice share the same conventional  frame  \{x$^{'}$, y$^{'}$, z$^{'}$\} (see Fig.\ref{Fig.1}(c)). And the D$_{3d}$ symmetry can be expressed by introducing a higher symmetric frame for the spin axis. The spin coordinate system changed to  \{x, y, z\} (see Fig.\ref{Fig.1}(c)) after corresponding rotation. In the new coordinate system, \{x$^{'}$, y$^{'}$, z$^{'}$\} unit vector can be expressed as $\hat{x}^{'}={1\over\sqrt 2}(1,-1,0), \hat{y}^{'}={1\over\sqrt{6}}(1,1,-2), \hat{z}^{'}={1\over\sqrt3}(1,1,1)$. 

In other words, the spin components in Hamiltonian (\ref{Habc}) can be expressed as 
\begin{eqnarray*}
&&S^x = {\sqrt 2\over2}S^{'x}+{\sqrt 6\over6}S^{'y}+{\sqrt 3\over3}S^{'z},\\ 
&&S^y =  {-\sqrt 2\over2}S^{'x}+{\sqrt 6\over6}S^{'y}+{\sqrt 3\over3}S^{'z},\\
&& S^z = {-\sqrt 6\over3}S^{'y}+{\sqrt 3\over3}S^{'z}.\ 
\end{eqnarray*}

By introducing the transformed spin into Hamiltonian (\ref{Habc}), we can get the spin Hamiltonian model in conventional  frame \{x$^{'}$, y$^{'}$, z$^{'}$\}  
\begin{eqnarray*}\label{Habcd}
H_x=\sum_{\langle i,j=i+a_1\rangle} &&(J+\frac{K}{2}-\Gamma^{'})S^{'x}_iS^{'x}_j \\
&&+(J+\frac{K}{6}-\frac{2}{3}\Gamma-\frac{1}{3}\Gamma^{'})S^{'y}_iS^{'y}_j\\
&&+(J+\frac{K}{3}+\frac{2}{3}\Gamma+\frac{4}{3}\Gamma^{'})S^{'z}_iS^{'z}_j\\
&&+(\frac{\sqrt 3}{6}K+\frac{\sqrt 3}{3}\Gamma-\frac{\sqrt 3}{3}\Gamma^{'})(S^{'x}_iS^{'y}_j+S^{'y}_iS^{'x}_j)\\
&&+(\frac{\sqrt 6}{6}K-\frac{\sqrt 6}{6}\Gamma+\frac{\sqrt 6}{6}\Gamma^{'})(S^{'x}_iS^{'z}_j+S^{'z}_iS^{'x}_j)\\
&&+(\frac{\sqrt 2}{6}K-\frac{\sqrt 2}{6}\Gamma+\frac{\sqrt 2}{6}\Gamma^{'})(S^{'y}_iS^{'z}_j+S^{'z}_iS^{'y}_j),\
\end{eqnarray*}

\begin{eqnarray*}\label{Habcd}
H_y=\sum_{\langle i,j=i+a_2\rangle} &&(J+\frac{K}{2}-\Gamma^{'})S^{'x}_iS^{'x}_j \\
&&+(J+\frac{K}{6}-\frac{2}{3}\Gamma-\frac{1}{3}\Gamma^{'})S^{'y}_iS^{'y}_j\\
&&+(J+\frac{K}{3}+\frac{2}{3}\Gamma+\frac{4}{3}\Gamma^{'})S^{'z}_iS^{'z}_j\\
&&+(-\frac{\sqrt 3}{6}K-\frac{\sqrt 3}{3}\Gamma+\frac{\sqrt 3}{3}\Gamma^{'})(S^{'x}_iS^{'y}_j+S^{'y}_iS^{'x}_j)\\
&&+(-\frac{\sqrt 6}{6}K+\frac{\sqrt 6}{6}\Gamma-\frac{\sqrt 6}{6}\Gamma^{'})(S^{'x}_iS^{'z}_j+S^{'z}_iS^{'x}_j)\\
&&+(\frac{\sqrt 2}{6}K-\frac{\sqrt 2}{6}\Gamma-\frac{\sqrt 2}{6}\Gamma^{'})(S^{'y}_iS^{'z}_j+S^{'z}_iS^{'y}_j),\
\end{eqnarray*}

\begin{eqnarray*}\label{Habcd}
H_z=\sum_{\langle i,j=i+a_3\rangle} &&(J-\Gamma)S^{'x}_iS^{'x}_j \\
&&+(J+\frac{2}{3}K+\frac{1}{3}\Gamma-\frac{4}{3}\Gamma^{'})S^{'y}_iS^{'y}_j\\
&&+(J+\frac{K}{3}+\frac{2}{3}\Gamma+\frac{4}{3}\Gamma^{'})S^{'z}_iS^{'z}_j\\
&&+(-\frac{\sqrt 2}{3}K+\frac{\sqrt 2}{3}\Gamma-\frac{\sqrt 2}{3}\Gamma^{'})(S^{'y}_iS^{'z}_j+S^{'z}_iS^{'y}_j),\
\end{eqnarray*}
where $H_x$, $H_y$, $H_z$ is the Hamiltonian of the $x$-, $y$-, $z$-bond, respectively. Above Hamiltonian can take a compact form by a $90^\circ$ rotation, 
$\hat{x}^{'}\to \hat y^{'}={1\over\sqrt 6}(1,1,-2), \hat y^{'}\to-\hat{x}^{'}={1\over\sqrt{2}}(-1,1,0), \hat{z}^{'}={1\over\sqrt3}(1,1,1)$. Thus the effective model is transformed into\cite{10.21468/SciPostPhysCore.3.1.004},
\begin{equation}\label{Hxyz}
\begin{split}
H=\sum_{\langle i,j\rangle} &J_{zz}S^z_iS^z_j +J_\pm(S^+_iS^-_j+S^-_iS^+_j) \\ 
&+ J_{\pm \pm}(\gamma_{ij}S^+_iS^+_j+\gamma^*_{ij}S^-_iS^-_j) \\
&+J_{\pm z}\big(\gamma^*_{ij}S^+_iS^z_j+\gamma_{ij}S^-_iS^z_j+ \langle i \leftrightarrow j\rangle\big),
\end{split}
\end{equation}
where $S^\pm_i=S^x_i \pm i S ^y_i$, and the phase factors $\gamma_{ij}$ along the $a_1$, $a_2$, $a_3$ bonds (see Fig.\ref{Fig.1} (b)) are $e^{i2\pi/3}, e^{-i2\pi/3}, 1$ respectively.

\par
It has also mentioned in this paper that there exists DM interaction in YOCl. Structurally, the nearest neighbor bonds have inversion symmetry according to their bond centers, but the next nearest-neighbor bond has no central inversion symmetry, so DM interactions are allowed. There is a mirror symmetry plane perpendicular to the next nearest-neighbor bond, so the DM vector is perpendicular to the bond\cite{PhysRev.120.91}. Furthermore, the anion Cl$^-$ is far away from the next-nearest bonds and have weaker effects, so the next nearest-neighbor rare-earth ions main exchange throng the O$^{2-}$ ions. And the DM vector $\pmb D_{ij}$ is almost perpendicular to the triangular cation-layer plane. In the following discussion, we analyze the symmetry restraint to the direction of $\pmb D_{ij}$, where $(ij)$ is oriented in the counterclockwise manner.

Here we provide the orientation of the DM vector $\pmb D_{ij}$ in the new spin coordinate system \{x, y, z\} (see Fig.\ref{Fig.1}(c)). Firstly, we consider the component perpendicular to the Honeycomb plane. In the new frame, $\pmb c\parallel {1\over\sqrt3}(1,1,1)$, so, $\pmb D\parallel {1\over \sqrt3}(1,1,1)$. Furthermore, owing to the $D_{3d}$ symmetry, the $c$-component of the $\pmb D_{ij}$ vector on the two sublattices have the same sign. 


Then we analyze the in-plane components although they are relatively weak. Since the $\pmb D$ vector is perpendicular to the next nearest-neighbor bond\cite{PhysRev.120.91}, the allowed direction is one of $\pm {1\over\sqrt6}(2x-y-z)$, $\pm {1\over\sqrt6}(2y-z-x)$, $\pm {1\over\sqrt6}(2z-x-y)$, depending on the bond direction. 


\bibliography{ReOCl.bib}

\end{document}